\documentclass[reprint,amsmath,amssymb,aps,prl,nobibnotes,floatfix,superscriptaddress,parskip=never]{revtex4-1}

\usepackage{etoolbox}
\usepackage{graphicx}
\usepackage{xcolor}
\definecolor{UniBlue}{RGB}{46,48,146}
\usepackage[
    colorlinks=true,
    citecolor=UniBlue,
    linkcolor=UniBlue,
    urlcolor=UniBlue]{hyperref}
\usepackage{bm}

\makeatletter
\pretocmd{\NAT@open}{\begingroup\color{\@citecolor}}{}{}
\apptocmd{\NAT@close}{\endgroup}{}{}
\makeatother

\begin{document}
\title{Swirling due to misaligned perception-dependent motility}

\author{Rodrigo Saavedra}
 \email{r.saavedra@fz-juelich.de}
\author{Gerhard Gompper}
 \email{g.gompper@fz-juelich.de}
\author{Marisol Ripoll}
 \email{m.ripoll@fz-juelich.de}
 \affiliation{%
Theoretical Physics of Living Matter, 
Institute for Advanced Simulation,
Forschungszentrum J\"ulich, 52425 J\"ulich, Germany
}
\date{\today}

\begin{abstract}
A system of particles with motility variable in terms of a vision-type of perception 
is here investigated by a combination of Langevin dynamics simulations 
in two-dimensional systems and an analytical approach based on conservation law principles. 
Persistent swirling with  predetermined direction is here induced by differentiating 
the self-propulsion direction and the perception cone axis.
Clusters can have a fluid-like center with a rotating outer layer, or display a solid-like 
rotation driven by the outer layer activity. 
Discontinuous motility with misaligned perception might therefore serve as a powerful self-organization strategy in micro-robots.
\end{abstract}

\maketitle

Self-assembly into swirling cohesive groups is a frequent strategy used by living organisms in a wide range of length scales~\cite{bechinger2016active}. %
Performing circular trajectories around a common center has shown to increase the structure resistance to 
external perturbations, and it is used both with foraging optimization of predator protection purposes~\cite{brilliantov21}. %
At the macroscopic level, examples are schools of fish~\cite{tunstrom2013collective} or swarms of insects~\cite{attanasi2014collective}, 
and at the microscopic level, vortex formation in colonies of bacteria~\cite{czirok1996formation}. %
Artificially, swirling has been obtained by employing external magnetic fields to control colloidal micro-robots~\cite{xie2019reconfigurable} and nanoparticles~\cite{yu2018pattern}; by employing light to locally control Janus particles~\cite{khadka2018active,bauerle2020formation}; or by employing external electric fields to Quincke rollers in circular confinement~\cite{bricard2015emergent,vlahovska2,zhang_r1,zhang2021active,zhang2021persistence}. %
Most of the mechanisms for vortex formation involve intrinsic particle chirality~\cite{ref1,ref2,mecke2}, or alternatively
a combination of attractive forces, to ensure group formation, and interparticle alignment~\cite{vicsek2012collective}. %
Vortex formation has also  been found in systems with no explicit alignment, where agents actively turn towards a crowd~\cite{zhang2021active}, 
with an externally applied torque~\cite{williams2016transmission,dewangan2019rotating}, delayed attractions~\cite{chen2023active,wang2023spontaneous}, or sedimenting active droplets~\cite{maas22rot}. %
To find different and still simple strategies that result in a controllable vortex formation remains a challenge. 
This can find very interesting applications in the development of smart active materials, or self-organizing microrobots~\cite{yu2023programmable,roadmap20,ceron2023programmable,gardi2022microrobot,hou2023review,lecheval2023smart,xie2019reconfigurable}. 

Navigation strategies based on visual-type of perception are intrinsic to many living systems and result into a very rich variety of flocking behaviors, such as aggregation, milling, or~meandering~\cite{bagarti2019milling,bastien2020model,durve2018active,lan2021autonomous,lavergne2019group,moussaid2011how,newman2008effect,pearce2014role,strandburg-peshkin2013visual}. %
Visual-type of perception restricts the interactions to neighbors placed inside a finite cone with  propulsion direction as the symmetry axis, and %
tip at the particle position.
This limited field of interaction is common to most animals, and implies non-reciprocal interactions which have shown to lead to a rich collective behavior~\cite{durve2018active,saha2020scalar,knezevic2022collective,kreienkamp2022clustering,vitelli21}. %
Inspired by such biological systems, minimal microscopic models have for example shown to lead to gas-like, milling behavior, worm-like or aggregate structures~\cite{barberis2016largescale,peruani2017hydrodynamic,costanzo2018,costanzo2019,negi2022,loos2023longrange,stengele2022group,fruchart2021nonreciprocal}. %
Provided a particular perception, a rule is required to determine the particle action, which ultimately governs the system properties. %
To apply this concept to the design of synthetic materials, a simple, yet interesting strategy is a discontinuous self-propulsion, which can be implemented as a switch 
in experiments with external laser heating~\cite{bauerle2018selforganization,bauerle2020formation,soker2021how}, 
and has been employed in systems of particles with quorum sensing~\cite{bauerle2018selforganization,fischer2020quorumsensing}. %
A computer assisted feedback loop to control the individual interactions of synthetic colloids has shown to form cohesive groups~\cite{lavergne2019group,chen2022collective}, 
and if the interaction accounts for the perception of the neighbors positions, this can lead to a spontanous breakdown of symmetry and to the formation of swirls~\cite{bauerle2020formation}.
Finally, misalignment between motion and interactions has been studied for single spinning colloids in a viscoelastic media, and micron-sized algae in confinement~\cite{narinder18,cao2023memoryinduced,bentley2022phenotyping}. 

\begin{figure}[b]
    \centering%
    \includegraphics[width=\linewidth]{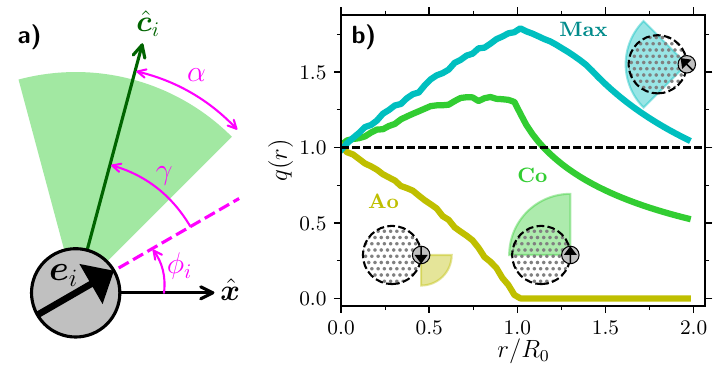}%
    \caption{\label{fig:1}%
        a)~Illustration of an active particle with misaligned visual perception. 
          The propulsion direction occurs in the $\bm{e}_i$ direction, the vision cone is centered in the direction $\bm{c}_i$ with half-width-angle~$\alpha$. 
          The relative angle between $\bm{e}_i$ and $\bm{c}_i$ is given by the misalignment $\gamma$. %
        b)~Perception radial profiles and sketches for co-oriented (Co), anti-oriented (Ao), and maximum perception (Max) oriented test particles, 
               with $r$ the distance to the cluster center of mass, in the initial homogeneous conditions, for $\gamma=\pi/4$. 
         }%
\end{figure}
In this letter, perception and motion are not considered to have the same direction, see Fig.~\ref{fig:1}.
Such misalignment shows to be sufficient to induce the formation of cohesive clusters with a predetermined rotation direction. %
Only the position of the neighbouring particles is relevant and not their orientations, which clearly differs from and simplifies existing strategies.
Cluster formation and rotation are investigated by simulations as a function of the misalignment angle $\gamma$ and the perception threshold $q^*$. 
These two quantities qualitatively modify the cluster rotation  and structure, which can be dilute or compact, homogeneous or non-homogeneous.    
These behaviors are quantitatively well-described by an analytical approach based on conservation principles.

A system of $N$ particles is here considered, each characterized by their position $\bm{r}_i$, 
and the orientation along which the propulsion takes place, $\bm{e}_i\equiv(\cos\phi_i,\sin\phi_i)^T$, 
with $\phi_i$ the angle between the orientation and the $x$ axis (see Fig.~\ref{fig:1}a). %
Each particle perceives its surrounding via the function,
\begin{equation}
    P_i = \sum_{j\in{c_i}}\frac{1}{r_{ij}},\quad {\rm if} \ r_{ij}< r_c,
    \label{eq:perception_function}
\end{equation}
for particles $j$ in the \emph{perception cone} $c_i$ of particle $i$, with $i,j=1,\ldots,N$, as illustrated in Fig.~\ref{fig:1}a. 
The perception is assumed to decay with the inverse of the interparticle distance $r_{ij}$, with  $r_c$ the maximum perception distance. 
The cone half-width angle is $\alpha$, while $\gamma$ is the angle displacement of the cone's symmetry axis with respect to the particle self-propulsion direction $\bm{e}_i$.
Interactions are non-reciprocal for $\alpha<\pi$, i.e.~a particle $j$ can be located inside the field of vision of another particle $i$, whereas $i$ is located outside the field of vision of $j$, %
such that the perception of any particle strongly depends on the orientation of its vision cone.

To complete the description, we define the normalized perceptions as $q=P/P_0$. %
The \emph{homogeneous perception}, $P_0$, is the perception of a particle placed in the center of a circular region with homogeneous density and radius $R_0=r_c/2$.  %
The homogeneous perception is expected to increase linearly with the width of the vision cone~$\alpha$, the system number density~$\rho=N/(\pi R_0^2)$, and the region radius, such that~$P_0=\alpha \rho R_0$ which, by construction, does not depend on $\gamma$.
Without misalignment, $\gamma=0$, the particle vision cone's symmetry axis and self-propulsion orientation coincide, and particles oriented towards or against  the center have a perception imbalance, which translates into the cluster coherence. %
With misalignment, $\gamma\neq{0}$,  not only the radial, but also the tangential orientation of the vision cone is relevant, and an imbalance appears also between Co-oriented (Co) and anti-oriented (Ao) particles. %
A Co-particle has a self-propulsion orientation tangential to the group's center, and its vision cone director $\bm{c}_i$ points towards the group center, see Fig.~\ref{fig:1}b. %
Conversely, an Ao-particle has the vision cone director pointing against the group center.
Perception profiles in Fig.~\ref{fig:1}b are numerically calculated for three test particles with fix orientations placed at different positions relative to the cluster center in a given initial configuration, with all other particles positions randomly chosen. 
Results are provided as an average over $10$ independent initial configurations. 
The maximum perception here corresponds to an oblique modification of the Co-oriented case.

\begin{figure*}[t]
    \centering
    \includegraphics[width=\linewidth]{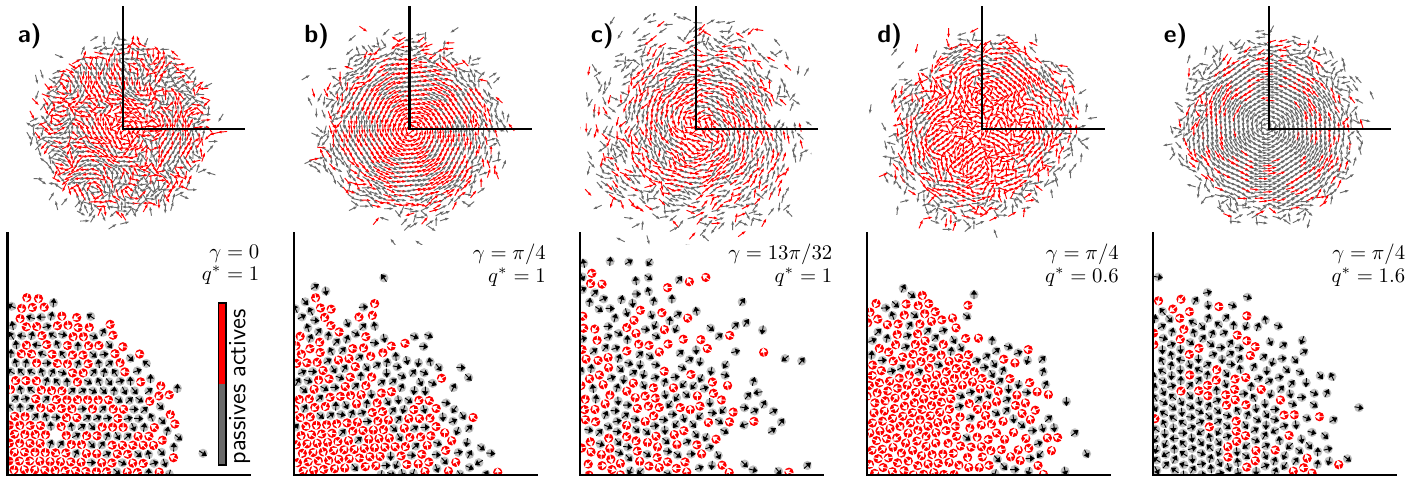}%
    \caption{\label{fig:2}%
        Snapshots of the system in stationary state for various $\gamma$ and $q^*$, with particles color-coded activity (see related movies in Ref.~\cite{sm}).  %
        The arrows in the upper row indicate average short-time displacements, illustrating cluster dynamics,  
        and axis-lines marks the area in the bottom row.  %
        The arrows in the bottom row correspond to particle orientations, illustrating the chosen bias. 
        Cohesive clusters showing behaviors:
        a)~compact non-swirling (no misalignment); %
        b)~solid-body rotation (misalignment); %
        c)~non-compact swirling (high misalignment); %
        d)~active core and swirling outer layer (low threshold); %
        e)~solid-body rotation, passive core with active dragging particles (high threshold).
    }%
\end{figure*}
The particles motion is governed by overdamped Langevin dynamics, which corresponds to micrometer-size Brownian particles 
\begin{equation}
    \begin{aligned}
        \dot{\bm{r}}_i &= v_i\bm{e}_i + \bm{f}^\text{EV}_i+ \sqrt{2 D_t}\bm{\xi}_i, \\
        \dot{\phi}_i &= \sqrt{2 D_r} \eta_i,
    \end{aligned}%
    \label{eqs:overdamped_langevin}%
 \end{equation}%
where, $\bm{\xi}_i$ and $\eta_i$ are translational and orientational white noises, and $D_t$ and $D_r$ the translational and rotational diffusion coefficients. %
The excluded volume force is~\mbox{$\bm{f}^\text{EV}=-\nabla{U}$}, with $U(\sigma,\epsilon)$ the Weeks-Chandler-Anderson potential, with $\sigma$ the particle diameter, and $\epsilon=100k_\mathrm{B}T$ the repulsion strength. %
The particles propulsion velocity $v_i$ is defined by a two-stage velocity~$v_i=v_0\Theta(q_i - q^*)$, with~$v_0$ a constant self-propulsion velocity, and~$q^*$ the normalized threshold perception value, similar to previous experiments~\cite{lavergne2019group}. %
Simulations start from a circular homogeneous configuration with radius $R_0$, and Eq.~(\ref{eqs:overdamped_langevin}) is integrated with the Euler algorithm and $\Delta t=10^{-5}$. %
Default parameters~\cite{lavergne2019group} fix the P{\'e}clet number $Pe=v_0/(\sigma D_r)$, as $Pe=4.8$.
All quantities are normalized or expressed in simulation units, here $\sigma$ and $D_t$. %
We present here a study of systems with $N=1000$ particles, $v_0=40$, $D_r= 8.3$, $\rho_0=0.51$, $k_\mathrm{B}T=1$, and $\alpha=\pi/4$. 
Other values of  $N$ or $r_c$ can be also chosen, considering these quantities modify the 
 homogeneous perception $P_0$, and some details of the results.

Typical snapshots in Fig.~\ref{fig:2} show particle positions, orientations, activities, and velocities calculated from the displacements during a time interval $\delta{t}=0.2$. %
The isotropic displacements in~Fig.~\ref{fig:2}a proves the lack of net motion in the absence of misalignment. %
Conversely, cases with misalignment $\gamma\neq{0}$ in~Fig.~\ref{fig:2}b-e clearly rotate. %
The misalignment mechanism selects as active the Co-particles, these are oriented with the vision cone inducing a bias in the motion which 
results in a net angular momentum. The torque arises then by this bias, without any additional orientation of the individual particles, 
such that the self-propulsion orientation field is isotropic in all cases, as shown in the zoom-ins in~Fig.~\ref{fig:2}. 
The distribution of active and passive particles in the cluster is mostly homogeneous for the $q^*=1$, see~Fig.~\ref{fig:2}a-c. 
This can be understood in terms of the perceptions profiles in Fig.~\ref{fig:1}b, 
since for  $q^*=1$ all Co-particles have perceptions above the threshold, and Ao-particle below. 
For lower values of $q^*$, the cluster is fluid-like and composed mainly of active particles, 
with passives only in the outer layer, see~Fig.~\ref{fig:2}d. 
This occurs because in the center all Co- and Ao-particles have perceptions above the threshold (see Fig.~\ref{fig:1}b), 
and only towards the cluster surface Ao-particles become passive.  
For higher $q^*$, the cluster is solid-like and composed largely of passive particles. 
In this case only Co-particles far from center have large enough perceptions to become active, 
and these ensure the cluster cohesion and rotation, see~Fig.~\ref{fig:2}e. 
Finally note that in all cases, passive particles outside the cluster will eventually reorient and become active rejoining the cluster. 
 
\begin{figure}[t]
    \centering%
    \includegraphics[width=\linewidth]{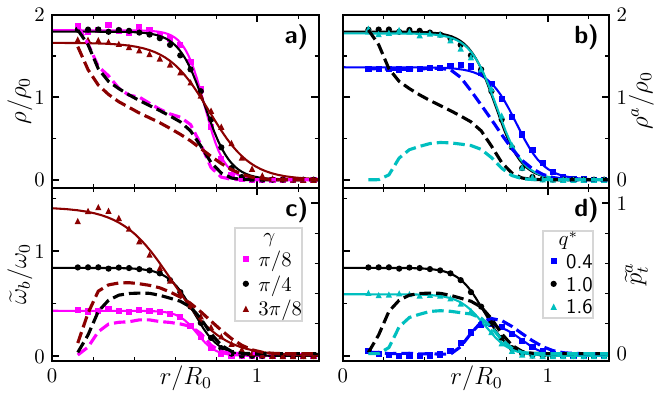}
    \caption{\label{fig:3}%
        Steady-state radial profiles for 
        a,b)~the number density of all particles $\rho$ (symbols), 
               and of only active particles $\rho^\mathrm{a}$ (dashed lines); %
        c,d)~angular velocity $\widetilde{\omega}/\omega_0$ (solid lines, left axes), 
              and tangential orientations of the active particles $\widetilde{p}_t^\mathrm{a}$ (dashed lines, right axes). %
        In a,c)~simulation results for fixed perception threshold $q^*=1$ and varying the misalignment $\gamma$; %
        In b,d) for fixed $\gamma=\pi/4$ and varying $q^*$. %
        The continuous lines are fits for the simulation data of $\rho$ and $\widetilde{\omega}$ with Eq.~(\ref{eq:tanh}) which determines the steady-state values $R_c$, $\rho_b$, and  $\omega_b$.
    }%
\end{figure}

A quantitative characterization of the cluster structure is performed via radial profiles of the number density calculated from the cluster's center of mass~$\bm{r}_\mathrm{cm}$, 
and separately for the density of active,  $\rho^\mathrm{a}$, and all particles, $\rho$. %
For $q^*=1$,~Fig.~\ref{fig:3}a shows that the density $\rho$ is constant at the center and decays at the interface, indication of a coherent cluster, 
while the active particles density shows an additional increase at the center. Both densities do not significantly change for $\gamma\leq\pi/4$,  which means 
that for too large misalignments, the cluster coherence is not affected.
With further increase of the misalignment, such as $\gamma=3\pi/8$ in~Fig.~\ref{fig:3}a, cluster size and interface width become slightly larger, while the number of active particles decreases.
In the limit of $\gamma\simeq\pi/2$, only particles moving tangentially to the cluster become active, such that no cluster coherence is possible. %
Given a fixed value of $\gamma$, the perception threshold variation does not affect much the overall cluster size (see Fig.~\ref{fig:3}b), but the motility distribution, as already explained for snapshots in~Fig.~\ref{fig:2}b,d,e.

The average radial profiles of the angular velocity, $\omega$, and of the active particles tangential polarization, $p_t^\mathrm{a}$, are shown in~Fig.~\ref{fig:3}c,d. 
These are normalized with the local relative density, $\widetilde{\omega}(r)\equiv \omega(r)\rho(r)/\rho_0$, to diminish noise in the cluster surface.
The average angular velocity is further normalized with $\omega_0=v_0/R_0$, the angular velocity of a Co-oriented particle placed at the initial cluster boundary.
Increasing $\gamma$ enhances the motion bias, which enlarges the average tangential orientation, 
and consequently also the induced torque and overall angular velocity, as shown in~Fig.~\ref{fig:3}c. 
In contrast, changing the perception threshold does not have a monotonic effect in the orientation and rotation properties.
For small values, such as $q^*=0.4$, the orientation of the active particles and rotation appears only in the outer layer. 
Particles in the cluster center perceive enough neighbours to become active independently of their orientation, 
such that this part remains fluid-like and both angular velocity and tangential active orientation have close to zero values. %
Closer to the cluster surface, only particles with a vision cone oriented towards the center become active, 
such that not only cohesion, but also a well-defined torque is induced. %
For increasing $q^*$ values, the selection of inwards oriented particles is more restrictive, making cohesion and compaction stronger.
Due to steric interactions in the compact state, active particles drag along the passive ones, resulting in 
solid-like rotations with a constant velocity at the cluster center, decaying towards the cluster boundary. 
For $q^*=1$, the distribution of active and passive particles is mostly homogeneous in the whole cluster, and  
both rotation and orientation are constant at the cluster center, decaying only at the cluster boundary. 
For larger values of  $q^*$, such as $q^*=1.6$,
density, orientation, and rotation of the active particles vanishes at the cluster center and become significant mostly for Max-oriented particles (see Fig.~\ref{fig:1}b) in the outside layer. 
The external active particles drag the rotation of the compact cluster, such that the angular velocity is constant in the center.   

\begin{figure}[h!]
    \centering%
    \includegraphics[width=\linewidth]{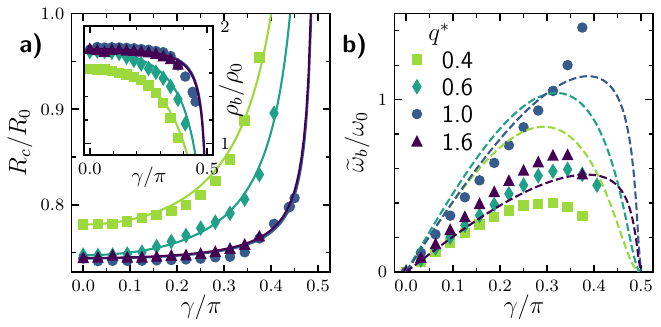}%
    \caption{\label{fig:4}%
        Cluster properties as a function of the misalignment $\gamma$, for $q^*=1$. %
        a)~Radius $R_c$, and  density~$\rho_b$ in the inset; 
        b)~angular velocity~$\omega_{b}$.
        Symbols are obtained as fits to the simulation results in Fig.~\ref{fig:3} with Eq.~(\ref{eq:tanh}); 
        continuous lines correspond to the analytical predictions in Eq.~(\ref{eq:Rc_gamma3}) for $R_c$, 
        related to it and Eq.~(S3) for $\rho_b$,  
        and in Eq.~(S14) for $\omega_{b}$ (see SM in ~\cite{sm}). %
    }%
\end{figure}

The radial profiles in Fig.~\ref{fig:3} mostly show to be constant at the cluster center with a soft decay at the cluster boundary,  
behavior which can be characterized by
\begin{equation}\label{eq:tanh}
\rho(r) = \frac{\rho_b}{2} \left[ 1 + \tanh\left(\frac{R_c-r}{2 \zeta}\right)\right], 
\end{equation}
where $\rho_b$ is the bulk density, $R_c$  the cluster radius, and $\zeta$ can be understood as the half-width of the cluster interface.  %
The values of $\rho_b$ and $R_c$ obtained from the fits are shown in Fig.~\ref{fig:4}a as a function of $\gamma$ for various values of $q^*$.   %
The radius $R_c$ in~Fig.~\ref{fig:4}a remains reasonably constant for a wide range of $\gamma$, and only for the largest misalignment
the radius rapidly increases until the cluster eventually dissolves becoming a gas. 

The functional form of $R_c(\gamma)$ can be analytically estimated by considering that (for details see SM~\cite{sm}): 
i)~the cluster is circular and has a stable size;
ii)~at the cluster surface, the number of particles diffusely leaving the cluster and actively joining the cluster exactly balances; 
iii)~without misalignment,  the cluster is cohesive with a size $R_{\gamma 0}$ determined by steric interactions,
this value is measured in simulations and tends to a close package configuration for moderately large values of $q^*$;
iv)~diffusion is given by $D_\mathrm{eff}$, the enhanced coefficient related to the average activity,
 which is then the only fitting parameter (see Fig.~S3). 
From the second condition it follows that $D_\mathrm{eff} \rho_b/R_c = v_op_r^a$, with $p_r^a$ the active particles radial polarization.
Such polarization can be calculated from the particles at the cluster boundary with orientations for which the vision cone encloses a portion of 
the cluster sufficiently large for  $q$ to excess $q^*$, which implies $p_r^a\sim\rho_b\cos\gamma$, (see Fig. S2 and Eqs. S9-S10).
With this we obtain
\begin{equation}
    R_c(q^*,\gamma) = \frac{D_\mathrm{eff}(q^*)}{A(q^*)}\left(\frac{1}{\cos\gamma} - 1\right) + R_{\gamma 0}(q^*), 
    \label{eq:Rc_gamma3}
\end{equation}
where $A(q^*)$ is related to the angle-activity threshold (see Fig.~S1 and S2~\cite{sm}). %
Figure~\ref{fig:4}a shows an excellent quantitative agreement of Eq.~(\ref{eq:Rc_gamma3}) with the simulation data.  
The measured cluster density $\rho_b$ shown in~the inset of~Fig.~\ref{fig:4}a can be directly related to the cluster 
size by $\rho_b=N/\pi{R_c^2}$, and the match of Eq.~(\ref{eq:Rc_gamma3}) and simulations results is also excellent. 
Similar comparison and agreement of $R_c$ and $\rho_b$ as a function of $q^*$ for fixed values of $\gamma$ is shown in Fig.~S4~\cite{sm}. %

The angular velocity radial profiles $\omega(r)$ in Fig.~\ref{fig:3}c,d can also be described by the functional 
form in Eq.~(\ref{eq:tanh}), by considering that the prefactor corresponds now to the bulk angular velocity $\omega_b$. 
For small $q^*$ cases, where the center is not rotating, Eq.~(\ref{eq:tanh}) applies only in the outer layer. 
The dependence of $\omega_b$ on $\gamma$ and $q^*$, as obtained in the fits, is shown in~Fig.~\ref{fig:4}b.
For small $\gamma$ values, $\omega_{b}$ grows linearly with $\gamma$ up to a maximum value from which it 
decreases until it vanishes for the dissolving cluster, for $\gamma \lesssim \pi/2$. %
In the case of $q^*\simeq 1$ and large $\gamma$ the rotation growth is faster, which is related to the fast central rotation, where the velocity is measured,
but it will eventually decrease again for $\gamma\to\pi/2$.  
The non-monotonous dependence of $\omega_\mathrm{max}$ with $q^*$ can be seen in Fig.~\ref{fig:4}b, and for fixed values of $\gamma$ also in Fig.~S4~\cite{sm}. %
For small $q^*$ values, almost all particles are active, such that the orientation bias and the rotation are small.
Increasing $q^*$ makes the orientation bias more pronounced
which increases the overall rotation speed, as long as the center of the cluster remains active. 
When the cluster center becomes passive and only the outside boundary remains active, the overall rotation speed decreases with $q^*$. %
The functional form of $\omega_b(\gamma)$ can be estimated by 
considering that the rotation is determined by the
active particles in the cluster surface $\omega_b = v_0{p}^a_t/R_c$, 
and then calculating ${p}^a_t$,  the tangential polarization of the active particles
following a similar procedure as before~(see details in~\cite{sm}). 
Figure~\ref{fig:4}b shows good
agreement between analytical predictions and simulation results for the cases corresponding to a compact rotating cluster, 
i.e. large $q^*$ and not too large $\gamma$, in particular the linear increase with $\gamma$.  
For more dilute cases, the estimations are only qualitative due to the performed approximation of the 
polarization at the cluster radius. 

In conclusion, the misalignment between the particles self-propulsion and perception cone directions is 
an effective mechanism to induce rotation in systems of agents with perception-dependent motility, while keeping their cohesion.
This is therefore a new and simple alternative method to induce swirling, 
also conceptually interesting since, in contrast to previous approaches~\cite{bauerle2020formation,barberis2016largescale,costanzo2018,costanzo2019}, 
it does not require explicit alignment,  nor external torques,  and the rotational direction is controlled.  
Our main conclusions can be extended to a larger range of parameters, in particular to systems a different number of particles $N$, 
width of the vision cone $\alpha$, or with a perception range $r_c$ considerably smaller than the cluster diameter (see one example case in SM~\cite{sm}).
The cluster properties and morphology can be tuned via these parameters, together with the perception threshold $q^*$ and the degree of misalignment $\gamma$. %
The agreement of theory and simulations supports the arguments that the cluster size is given with steric interations 
and the balance of particles diffusely leaving the cluster and particles actively joining the cluster together, 
and that the swirling velocity is determined by the tangential polarization of the particles selected as active at the cluster 
The proposed interaction mechanism is asymmetric and non-reciprocal and can be implemented in experiments of colloids activated by light or in robot swarms,
serving then as a self-organization strategy with various potential purposes.

\begin{acknowledgments}
This work was financially supported by the \hbox{CONACYT-DAAD} scholarship program. %
The authors gratefully acknowledge the computing time granted through JARA on the supercomputer JURECA~\cite{jureca21} at Forschungszentrum J{\"u}lich.
\end{acknowledgments}


%

\end{document}